\documentstyle[12pt,psfig,epsfig,axodraw]{article}
\advance\textheight by \topskip
\begin{document}
\tolerance=100000
\thispagestyle{empty}
\setcounter{page}{0}

\def\cO#1{{\cal{O}}\left(#1\right)}
\newcommand{\be}{\begin{equation}}
\newcommand{\ee}{\end{equation}}
\newcommand{\br}{\begin{eqnarray}}
\newcommand{\er}{\end{eqnarray}}
\newcommand{\ba}{\begin{array}}
\newcommand{\ea}{\end{array}}
\newcommand{\bi}{\begin{itemize}}
\newcommand{\ei}{\end{itemize}}
\newcommand{\bn}{\begin{enumerate}}
\newcommand{\en}{\end{enumerate}}
\newcommand{\bc}{\begin{center}}
\newcommand{\ec}{\end{center}}
\newcommand{\ul}{\underline}
\newcommand{\ol}{\overline}
\newcommand{\ra}{\rightarrow}
\newcommand{\sm}{${\cal {SM}}$}
\newcommand{\as}{\alpha_s}
\newcommand{\aem}{\alpha_{em}}
\newcommand{\ycut}{y_{\mathrm{cut}}}
\newcommand{\susy}{{{SUSY}}}
\newcommand{\Dir}{\kern -6.4pt\Big{/}}
\newcommand{\Dirin}{\kern -10.4pt\Big{/}\kern 4.4pt}
\newcommand{\DDir}{\kern -10.6pt\Big{/}}
\newcommand{\DGir}{\kern -6.0pt\Big{/}}
\def\Ecm{\ifmmode{E_{\mathrm{cm}}}\else{$E_{\mathrm{cm}}$}\fi}
\def\gluino{\ifmmode{\mathaccent"7E g}\else{$\mathaccent"7E g$}\fi}
\def\photino{\ifmmode{\mathaccent"7E \gamma}\else{$\mathaccent"7E \gamma$}\fi}
\def\mgluino{\ifmmode{m_{\mathaccent"7E g}}
             \else{$m_{\mathaccent"7E g}$}\fi}
\def\taugluino{\ifmmode{\tau_{\mathaccent"7E g}}
             \else{$\tau_{\mathaccent"7E g}$}\fi}
\def\mphotino{\ifmmode{m_{\mathaccent"7E \gamma}}
             \else{$m_{\mathaccent"7E \gamma}$}\fi}
\def\ML{\ifmmode{{\mathaccent"7E M}_L}
             \else{${\mathaccent"7E M}_L$}\fi}
\def\MR{\ifmmode{{\mathaccent"7E M}_R}
             \else{${\mathaccent"7E M}_R$}\fi}
\def\lsim{\buildrel{\scriptscriptstyle <}\over{\scriptscriptstyle\sim}}
\def\gsim{\buildrel{\scriptscriptstyle >}\over{\scriptscriptstyle\sim}}
\def\jp #1 #2 #3 {{J.~Phys.} {#1} (#2) #3}
\def\pl #1 #2 #3 {{Phys.~Lett.} {#1} (#2) #3}
\def\np #1 #2 #3 {{Nucl.~Phys.} {#1} (#2) #3}
\def\zp #1 #2 #3 {{Z.~Phys.} {#1} (#2) #3}
\def\pr #1 #2 #3 {{Phys.~Rev.} {#1} (#2) #3}
\def\prep #1 #2 #3 {{Phys.~Rep.} {#1} (#2) #3}
\def\prl #1 #2 #3 {{Phys.~Rev.~Lett.} {#1} (#2) #3}
\def\mpl #1 #2 #3 {{Mod.~Phys.~Lett.} {#1} (#2) #3}
\def\rmp #1 #2 #3 {{Rev. Mod. Phys.} {#1} (#2) #3}
\def\sjnp #1 #2 #3 {{Sov. J. Nucl. Phys.} {#1} (#2) #3}
\def\cpc #1 #2 #3 {{Comp. Phys. Comm.} {#1} (#2) #3}
\def\xx #1 #2 #3 {{#1}, (#2) #3}
\def\NP(#1,#2,#3){Nucl.\ Phys.\ \issue(#1,#2,#3)}
\def\PL(#1,#2,#3){Phys.\ Lett.\ \issue(#1,#2,#3)}
\def\PRD(#1,#2,#3){Phys.\ Rev.\ D \issue(#1,#2,#3)}
\def\preprint{{preprint}}
\def\Ord{\lower .7ex\hbox{$\;\stackrel{\textstyle <}{\sim}\;$}}
\def\OOrd{\lower .7ex\hbox{$\;\stackrel{\textstyle >}{\sim}\;$}}
\def\MCH {$\tilde\chi_1^+$}
\def \CH{{\tilde\chi}^{\pm}}
\def \LSP{\tilde\chi_1^0}
\def \SNU{\tilde{\nu}}
\def \BARSNU{\tilde{\bar{\nu}}}
\def \MLSP{m_{{\tilde\chi_1}^0}}
\def \MCH{m_{{\tilde\chi}^{\pm}}}
\def \MCHMIN {\MCH^{min}}
\def \ET{\not\!\!{E_T}}
\def \LL{\tilde{l}_L}
\def \LR{\tilde{l}_R}
\def \MLL{m_{\tilde{l}_L}}
\def \MLR{m_{\tilde{l}_R}}
\def \MSNU{m_{\tilde{\nu}}}
\def \PROCESS{e^+e^- \rightarrow \tilde{\chi}^+ \tilde{\chi}^- \gamma}
\def \PI{{\pi^{\pm}}}
\def \DM{{\Delta{m}}}
\newcommand{\bQ}{\overline{Q}}
\newcommand{\ad}{\dot{\alpha }}
\newcommand{\bd}{\dot{\beta }}
\newcommand{\dd}{\dot{\delta }}
\def \CH{{\tilde\chi}^{\pm}}
\def \MCH{m_{{\tilde\chi}_1^{\pm}}}
\def \LSP{\tilde\chi_1^0}
\def \MUL{m_{\tilde{u}_L}}
\def \MUR{m_{\tilde{u}_R}}
\def \MDL{m_{\tilde{d}_L}}
\def \MDR{m_{\tilde{d}_R}}
\def \MSNU{m_{\tilde{\nu}}}
\def \MTAUL{m_{\tilde{\tau}_L}}
\def \MTAUR{m_{\tilde{\tau}_R}}
\def \mhf{m_{1/2}}
\def \MST{m_{\tilde t_1}}
\def \RPVC{\lambda'}
\def\tth{\tilde{t}\tilde{t}h}
\def\qqh{\tilde{q}_i \tilde{q}_i h}
\def\t1{\tilde t_1}
\def \ta1{\tilde\tau_1}
\def \pt{p{\!\!\!/}_T}  
\def\lapp{\mathrel{\rlap{\raise.5ex\hbox{$<$}}
                    {\lower.5ex\hbox{$\sim$}}}}
\def\gapp{\mathrel{\rlap{\raise.5ex\hbox{$>$}}
                    {\lower.5ex\hbox{$\sim$}}}}
\begin{flushright}
{CERN/TH/2004-226}\\
IISc/CHEP/2004-14\\
\end{flushright}
\begin{center}
{\Large \bf
Using Tau Polarization to Discriminate between SUSY Models and Determine 
SUSY Parameters at ILC
}\\[1.00
cm]
\end{center}
\begin{center}
{\large R. M. Godbole$^a$, Monoranjan Guchait$^{b, c}$  
{and} D. P. Roy$^{b, c}$}\\[0.3 cm]
{\it 
$^a$Centre for High Energy Physics, Indian Institute of Science\\
Bangalore, 560012,India\\
\vspace{0.2cm}
$^b$Tata Institute of Fundamental Research\\ 
Homi Bhabha Road, Mumbai-400005, India.\\
\vspace{0.2cm}
$^c$Theory Division, Department of Physics\\
CERN, CH 1211 Geneva, Switzerland}
\end{center}

\vspace{2.cm}

\begin{abstract}
{\noindent\normalsize 
In many SUSY models the first SUSY signal in the proposed International Linear 
Collider is expected to come from the pair production of $\ta1$, followed 
by its decay into $\tau+$ LSP. We study a simple and robust method of measuring the
polarization of this $\tau$ in its 1-prong hadronic decay channel,
and show how it can be used to discriminate between SUSY models and to determine SUSY 
parameters. 

}
\end{abstract}

\vspace{2cm}
\hskip1.0cm
\newpage

\section*{Introduction}
\label{sec_intro}
It is widely recognized now that the LHC should provide unambiguous
signals of supersymmetry; but one needs the data from the proposed International Linear 
Collider(ILC)
to discriminate
between different SUSY models as well as for model independent
determination of the SUSY parameters~\cite{accomo}. One expects the lighter tau
slepton,

\be
\tilde\tau_1 = \tilde\tau_R \sin\theta_\tau + \tau_L \cos\theta_\tau,
\label{one}
\ee
to play a very important role in the SUSY study at ILC, since it is expected to be the
lightest sfermion and the next to lightest superparticles (NLSP) over a large range of SUSY
parameters in many important versions of the MSSM~\cite{kane,pdg}. Consequently the first 
SUSY signal at ILC is expected to come from the pair production of $\tilde\tau_1$.

In the universal SUGRA model the superparticle masses at the weak scale are related to 
the universal 
SUSY breaking mass parameters at the GUT scale, $m_0$ and $m_{1/2}$, by the 
renormalization group equations(RGE). The lighter neutralinos($\tilde Z_{1,2}$)
and chargino state($\tilde W_1$) are dominated by the U(1) and SU(2) gauginos, 
$\tilde B$ and $\tilde W$, with masses 

\br
M_1 = \frac{ 5 \alpha_1}{3 \alpha_2}M_2 \simeq 0.5 M_2, \nonumber \\
M_2 = \frac{\alpha_2}{\alpha_G} m_{1/2}\simeq 0.8 m_{1/2}. 
\label{eq:two}
\er

The left and right slepton masses are given by 
\br
\MLL^2 = m_0^2 + 0.5 m_{1/2}^2, \nonumber \\
\MLR^2 = m_0^2 + 0.15 m_{1/2}^2.
\label{eq:three}
\er
 
At moderate to large $\tan\beta$($\ge 5$) the ${\tilde\tau}_{L,R}$ masses are suppressed 
by negative 
contributions to the above RGE from the $\tau$ yukawa coupling,
\br
h_\tau^2 = \frac{g^2 m_\tau^2 }{2 m_W^2 \cos^2\beta} 
\label{eq:four}
\er
which is twice as large in $\tilde\tau_R$ as in $\tilde\tau_L$. Moreover there is 
significant mixing 
between the $\tilde\tau_{L,R}$ states, as represented by the off-diagonal term in the 
mass matrix  
\br
m^2_{LR} = -m_\tau(A_\tau + \mu \tan\beta) 
\label{eq:five}
\er
which suppresses the mass of the lighter eigenstate $\tilde\tau_1$ further down. 
Thus $\ta1$ is 
predicted to be 
the lightest sfermion over the moderate to large $\tan\beta$($\geq$ 5) region, 
which is favored by the LEP 
limit on the light Higgs boson mass{\footnote {Although there is no strict LEP 
limit on $\tan\beta$ for the 
maximal mixing scenario corresponding to $A_t \simeq \sqrt{6} m_{\tilde t} 
\simeq 2.5$ TeV, a moderate value of
$ A_t \le 1$ TeV implies small mixing and $\tan\beta \geq$ 5~\cite{pdg,hwg}.}. 
Moreover from eqs.~\ref{eq:two}-\ref{eq:five} one expects $\tilde\tau_1$ to be the NLSP 
after $\tilde Z_1$, 
over half the parameter space
\br
m_0 < m_{1/2}.
\label{eq:six}
\er
 The RGE of eqs.\ref{eq:two} and ~\ref{eq:three} may not hold in non-universal 
versions of SUGRA and the 
gauge or anomaly mediated supersymmetry breaking models (GMSB and AMSB). Nonethless 
the suppressions
coming from the Yukawa coupling and 
the off-diagonal mass term of eqs.~\ref{eq:four} and ~\ref{eq:five} continue to 
hold in these models for 
moderate to large values of $\tan\beta$. Consequently the $\ta1$ is expected to be the NLSP over important 
ranges of parameter space in these models as well. 
For simplicity we shall consider the CP conserving versions of these models. 

The superparticle spectrum for a set of benchmark points for the above mentioned 
SUSY models, 
called the Snowmass
points and slopes(SPS), are given in Ref.~\cite{snowmass}. They satisfy all the 
experimental constraints 
including the cosmological constraint on the SUSY dark matter (DM) relic density. Three 
of the five universal SUGRA points 
have $\ta1$ as the NLSP, including the typical SUGRA points(SPS1). The two exceptions 
represent the 
focus point region $m_0 \gg m_{1/2}$ and the so called funnel region($\tan\beta=55$
and $m_0 > m_{1/2}$). The DM constraint is satisfied through the pair annihilation 
of $\tilde Z_1$ by their
coupling to the Z boson via their higgsino components in the focus point region and 
by their coupling 
to the pseudoscalar Higgs boson in the funnel region. In the other cases it is 
satisfied through the pair annihilation of $\tilde Z_1$ via a relatively light $\ta1$ 
exchange $\tilde Z_1 \tilde Z_1 \ra \tau^+ \tau^-$ or its co-annihilation with a nearly 
dengenerate $\ta1$. The 
list contains a non-universal SUGRA point(SPS6), where the $\tilde Z_1$ has a significant higgsino 
component and a roughly similar mass as $\tilde Z_2$ and $\tilde W_1$. In this case $\ta1$ 
is slightly 
heavier than this cluster. It should 
be added here that specific non-universal SUGRA models have been considered, where the 
gaugino masses arise 
from the vacuum expectation value(vev) of a non-singlet chiral superfield, belonging to the 75 or 200 representations of the 
GUT SU(5) 
group~\cite{huitu}. In these models the LSP consists of a set of degenerate higgsino $\tilde Z_{1,2}$ and 
$\tilde W_1$. The $\ta1$ turns out to be the next heavier state over the 
region(\ref{eq:six}) for 
moderate 
to large $\tan\beta$. The list of Ref.\cite{snowmass} contains two GMSB points, where the $\ta1$ is 
the NLSP 
to a very light gravitino $\tilde G$ in one case and it is slightly heavier 
than the $\tilde Z_1$ in 
the other. 
Finally it contains a AMSB point, where the $\ta1$ is the NLSP to a degenerate pair 
of $\tilde Z_1, \tilde W_1$ LSP, representing the $\tilde W$.

One sees from the above discussion that the $\ta1$ is expected to be the NLSP in a wide 
class of SUSY models. 
Thus one expects an unambiguous
SUSY signal from the pair production process
\be
e^+e^- \rightarrow \tilde\tau^+_1 \tilde\tau^-_1,
\label{eq:seven}
\ee
followed by the decays
\be
\tilde\tau^\pm_1 \rightarrow \tau^\pm \tilde Z_1.
\label{eq:eight}
\ee

In particular one can have a right polarized electron beam $(P_e
= +1)$, which couples only to the $U(1)_Y$ gauge boson $B$.  Thus for
$s \gg m^2_Z$, $\sigma (\tilde\tau_R) = 4\sigma (\tilde\tau_L)$; so
that one can predict the $\ta1$ pair production cross section in terms of 
its mass and mixing angle~\cite{nojiri,boos}. The mass can be estimated from a 
threshold scan and the mixing angle from the production cross section. Hence 
one can then predict the polarization of $\tau$ coming from the 
decay (\ref{eq:eight}) in terms of the composition of the LSP, i.e.
\br
P_\tau &=& {\left(a^R_{11}\right)^2 - \left(a^L_{11}\right)^2 \over
\left(a^R_{11}\right)^2 + \left(a^L_{11}\right)^2}, \nonumber
\\[2mm]
a^R_{11} &=& - {2g \over \sqrt{2}} N_{11} \tan\theta_W
\sin\theta_\tau - {gm_\tau \over \sqrt{2} m_W \cos\beta} N_{13}
\cos\theta_\tau, \nonumber \\[2mm]
a^L_{11} &=& {g \over \sqrt{2}} \left[N_{12} + N_{11}
\tan\theta_W\right] \cos\theta_\tau - {gm_\tau \over \sqrt{2} m_W
\cos\beta} N_{13} \sin\theta_\tau,
\label{eq:nine}
\er
where
\be
\tilde Z_1 = N_{11} \tilde B + N_{12} \tilde W + N_{13} \tilde H_1 +
N_{14} \tilde H_2,
\label{eq:ten}
\ee
and we have made the collinear approximation($m_\tau \ll m_{\ta1}$).

In this work we investigate the prospect of using $\tau$ polarization to probe the composition of 
$\tilde Z_1$ at the ILC. We shall use a simple and powerful method of measuring $\tau$ polarization via 
its inclusive 
1-prong hadronic decay, which was suggested in ~\cite{bullock,dp} in the context of charged Higgs 
boson signal. 
The efficacy of this method for $H^\pm$ search has been corroborated by detailed simulation 
studies by both 
the CMS and ATLAS groups at LHC~\cite{ketevi}. More recently the method has been used in the context of the 
$\ta1$ signature at Tevatron and LHC~\cite{guchait}. However this seems to us to be the first 
application of 
this method
for a $e^+ e^-$ collider signal. It should be noted here that some of the $\tilde\tau$ signal studies at 
$e^+ e^-$ collider have used the effect of $\tau$ polarization in its exclusive decay channels 
$\tau \to \pi \nu$~\cite{nojiri,boos} and $\tau \to \rho \nu$~\cite{nojiri}. 
This has its own advantage if one wants e.g. to reconstruct the 
$\tilde\tau$ mass from its decay kinematics rather than via threshold scan. 
But as a method of simply measuring the $\tau$ polarization the inclusive 
channel has several advantages over the exclusive ones, since the latter 
suffers from lower statistics due to the branching fraction of the 
particular exclusive channel as well as larger systematic error due to the 
contaminations from the other decay channels. The estimation of this 
systematic error depends heavily on the ILC detector parameters, which is 
beyond the scope of this paper. We can only refer the interested reader to the 
paper of Nojiri et. al.~\cite{nojiri} who have done a Monte Carlo simulation
incorporating the parameters of a specific model detector. In particular one 
can see
from their fig.7 a significant contamination of the $\tau \rightarrow \rho \nu$
decay channel from $\tau \rightarrow a_1\nu$. On the other hand the method 
advocated here uses a feature of $\tau$ polarization, to which the 
$\pi$, $\rho$ and $a_1$ decay channels contribute coherently. This obviates 
the necessity to separate these exclusive decay channels from one another as 
we see below. In view of the expected improvements in a future ILC 
detector, it is quite possible that the systematic errors in the exclusive
analysis will be reduced. In such a situation, the information obtained 
from both the inclusive and the exclusive analysis, can be used effectively
together.

\section*{~$\tau$ Polarization}

The hadronic decay channel of $\tau$ is known to be sensitive to $\tau$
polarization~\cite{bullock,dp}.  We shall concentrate on the 1-prong hadronic decay of
$\tau$, which is best suited for $\tau$ identification~\cite{ketevi}.  It accounts
for 80\% of $\tau$ hadronic decay and 50\% of its total decay width.
The main contributors to the 1-prong hadronic decay are~\cite{pdg},
\[
\tau^\pm \rightarrow \pi^\pm \nu(12.5\%), \ \rho^\pm \nu(26\%), \
a^\pm_1 \nu (7.5\%),
\]
where the branching fractions for $\pi$ and $\rho$ include the small
$K$ and $K^\star$ contributions respectively, which have identical
polarization effects.  Together they account for 90\% of the 1-prong
hadronic decay.  The CM angular distribution of $\tau$ decay into
$\pi$ or a vector meson $v (= \rho, a_1)$ is simply related to its
polarization via
\br
{1 \over \Gamma_\pi} {d\Gamma_\pi \over d\cos\theta} &=&
{1\over2} (1 + P_\tau \cos\theta) \nonumber \\
{1 \over \Gamma_v} {d\Gamma_{v L,T} \over d\cos\theta} &=&
{{1\over2} m^2_\tau, m^2_v \over m^2_\tau + 2m^2_v} (1 \pm P_\tau
\cos\theta),
\label{eq:eleven}
\er
where $L,T$ denote the longitudinal and transverse states of the
vector meson.  The fraction $x$ of the $\tau$ lab momentum carried by
its decay meson is related in each case to $\theta$ in the collinear approximation via

\be
x = {1\over2} (1 + \cos\theta) + {m^2_{\pi,v} \over 2m^2_\tau} (1 -
\cos\theta).
\label{eq:twelve}
\ee
The only measurable $\tau$ momentum is the visible momentum of the
$\tau$-jet,
\be
p_{\tau-{\rm jet}} = x p_\tau.
\label{eq:thirteen}
\ee
It is clear from eqs.~\ref{eq:eleven}-\ref{eq:thirteen} that the hard $\tau$-jets 
are dominated
by $\pi,\rho_L$ and $a_{1L}$ for $P_\tau = +1$, while they are
dominated by $\rho_T$ and $a_{1T}$ for $P_\tau = -1$.  The two can be
distinguished by exploiting the fact that the transverse $\rho$ and
$a_1$ decays favor even sharing of momentum among the decay pions,
while the longitudinal $\rho$ and $a_1$ decays favor uneven sharing,
where the charged pion carries either very little or most of the
vector meson momentum.  Therefore the fraction of the visible $\tau$-jet
momentum carried by the charged prong,
\be
R = p_{\pi^\pm}/p_{\tau-{\rm jet}},
\label{eq:fourteen}
\ee
is predicted to be peaked at the two ends $R < 0.2$ and $R > 0.8$ for
$P_\tau = +1$, while it is peaked at the middle for $P_\tau = -1$~\cite{bullock,dp}.
Thus the $R$ distribution of the hard $\tau$-jet can be used
effectively to measure $P_\tau$,where the numerator can be measured from the tracker and the denominator 
from the calorimetric energy deposit of the $\tau$-jet.  We shall use a jet hardness cut of
\be
p^T_{\tau-{\rm jet}} > 25 \ {\rm GeV}, \cos\theta_{\tau-{\rm jet}} < 0.75 
\label{eq:fifteen}
\ee
for $\tau$-identification. But for the sake of comparison we shall show the results for a 
harder cut of  $p^T_{\tau-{\rm jet}} > 50$~GeV as well.
It is well known that this signal can be effectively separated from the standard model
background via modest cuts on the acoplanarity of the two $\tau$-jets and the 
missing-$p_T$\cite{nojiri}. However we shall not consider these cuts here, since they do not affect the 
measurement of $P_\tau$.

In our Monte Carlo simulation we shall use the {\tt TAUOLA} packages~\cite{tauola}
for polarized $\tau$ decay into inclusive 1-prong hadronic channel. In addition to 
the above 
mentioned resonances it includes the small nonresonant contribution to the 1-prong 
hadronic $\tau$ decay as well. 
There is some uncertainty in the parameterizations of the $\pi,\rho$, 
$a_1$ and the non-resonant contributions which have been tuned to the $\tau$ decay 
data of different 
experiments by the respective collaborations. Indeed we expect this to be the main source of uncertainty in 
estimating the $\tau$-polarization from the R-distribution. To estimate the rough size 
of this uncertainty we 
shall compare the results obtained with the parameterizations of the 
{\tt ALEPH} 
and the {\tt CLEO} 
collaborations~\cite{tauola}. We shall use the {\tt PYTHIA} event generator~\cite{pythia} 
to simulate the 
$\ta1$ pair production and decay(eqs.~\ref{eq:seven},~\ref{eq:eight}) at the ILC.

For illustrative purpose we shall show the $R$ distribution of hard
$\tau$-jets coming from (\ref{eq:seven}) and (\ref{eq:eight}) for
\be
\sqrt{s} = 350 \ {\rm GeV}, \ \ \ m_{\tilde\tau_1} = 150 \ {\rm GeV},
\ \ \ m_{\tilde Z_1} = 100 \ {\rm GeV},
\label{eq:sixteen}
\ee
and different values of $P_\tau$, corresponding to different SUSY
models. These masses are close to the $\ta1$ and the $\tilde Z_1$ masses of the 
typical SUGRA point(SPS1)
as well as the other relevant points of Ref.~\cite{snowmass}. The size of the signal cross
section for this beam energy and the $\tilde\tau_1$ mass is $\sim 200 fb$, while
the typical luminosity for ILC is $\sim$ 100$fb^{-1}$. Correspondingly we have done a Monte
Carlo simulation of 2$\times 10^4$ signal events.   

We shall consider three different strategies for using $P_\tau$ to
probe the SUSY model and determine the model parameters: (A)
Discrimination between SUSY models, (B) Partial determination of SUSY
parameters and (C) Complete determination of $EW$ SUSY parameters in
association with the measurements of chargino $(\tilde W_1)$
cross-section and mass.

\section*{A) Discrimination Between SUSY models:}

We consider four 
different SUSY models which are different versions of the MSSM -- i) Universal  Supergravity, 
ii) Higgsino LSP,
iii) Anomaly Mediated SUSY Breaking(AMSB), iv) Gauge Mediate SUSY Breaking(GMSB).

\begin{enumerate}
\item[{i)}] In the universal SUGRA model the LSP is dominated by the $\tilde B$
component $(N_{11})$.  Moreover at small $\tan\beta$ the
mixing angle $\cos\theta_\tau$ is small.  Thus we see from
eq.(\ref{eq:nine}) that
\be
P_\tau \simeq +1.
\label{eq:seventeen}
\ee
Since the mixing angle is determined by the off-diagonal term of eq.~\ref{eq:five},
large $\tan\beta$ corresponds to large mixing $(\cos\theta_\tau)$.
But there is an effective cancellation between the two terms of
$a^L_{11}$ in eq.(\ref{eq:nine}), so that $P_\tau$ remains very close to
$+1$ over practically the entire parameter space of interest~\cite{guchait}.
Indeed it was shown in the second paper of Ref.~\cite{guchait} that $P_\tau>$0.9 
throughout the allowed SUGRA parameter space, while $P_\tau >$0.95 in 
the $m_0 < m_{1/2}$ region of our interest. Thus 
eq.\ref{eq:seventeen} holds to a very good approximation at large $\tan\beta$ as well. 

\item[{ii)}] In nonuniversal SUGRA models, the gauge kinetic function and the resulting gaugino masses 
can be determined by a non-singlet chiral superfield, at the GUT
scale, belonging to the $SU(5)$ representations 24, 75 or 200.  For
the 75 and 200 representations the LSP is dominated by the Higgsino
component over most of the parameter space~\cite{huitu}.  Thus from
eq.(\ref{eq:nine})
\be
P_\tau \simeq \cos^2 \theta_\tau - \sin^2 \theta_\tau,
\label{eq:eighteen}
\ee
particularly at large $\tan\beta$, where Higgsino couplings are enhanced.
It should be mentioned here that in this case there is a nearly degenerate pair of 
light neutralinos $\tilde Z_{1,2}$, which are both dominated by the higgsino 
components. Thus the predicted $\tau$ polarizations of eq.~\ref{eq:eighteen} holds for 
each of them.  
\item[{iii)}] In AMSB model the LSP is dominated by the wino component
$(N_{12})$~\cite{randal}, i.e.
\be
P_\tau \simeq -1.
\label{eq:nineteen}
\ee

\item[{iv)}] Finally in GMSB the LSP is the gravitino $\tilde G$,
while the $\tilde\tau_1$ is expected to be the NLSP 
over a large part of parameter space~\cite{dicus}. In particular this is true for one of the two benchmark
points of GMSB in Ref.~\cite{snowmass}. 
Thus the $\tilde\tau_1 \rightarrow \tau
\tilde G$ decay implies
\be
P_\tau = \sin^2\theta_\tau - \cos^2\theta_\tau.
\label{eq:twenty}
\ee
It may be added here that for the other GMSB benchmark point of ~\cite{snowmass} the 
$\ta1$ is slightly heavier than a $\tilde B$ dominated $\tilde Z_1$, so that one expects essentially the 
same $P_\tau$ as in the universal SUGRA model.  
\end{enumerate}

Thus for $\cos\theta_\tau = 1/2$, which is a reasonable value of the mixing angle in the large $\tan\beta$
region, we have
\be
P_\tau (i,ii,iii,iv) \simeq +1, -1/2,-1,+1/2.
\label{eq:twenty1}
\ee
Fig.1 shows the resulting R distributions for the four $P_\tau$ values corresponding to 
$p^T_{\tau-jet} > 25$~GeV(upper pannel) and $p^T_{\tau-jet} > 50$~GeV(lower panel). The last bin 
near $R= 1$
represents the $\tau^\pm \ra \pi^\pm \nu$ contribution. Thus the R distribution is expected to peak at 
the two ends for $P_\tau>$0 states while it is expected to peak near the middle along with a depleted 
$\tau^\pm \ra \pi^\pm \nu$ contribution for  $P_\tau < $0. The difference increases with increasing
$p^T_{\tau-jet}$ cut as expected from eq.~\ref{eq:eleven}-\ref{eq:fourteen}. But even for the 
$p^T_{\tau-jet}>$25 GeV cut of eq.~\ref{eq:fifteen}, the differences in the R distribution are sufficient
to distinguish the four models from one another.

Fig.2 shows the fraction of events in the interval 0.2$ < R < $0.8,
\br
f=\frac{\sigma(0.2 < R< 0.8)}{\sigma_{tot}}
\label{eq:twenty2}
\er
against the $\tau$ polarization. This fraction is seen to decrease from 
65\% at $P_\tau=-$1 to 35\% at 
$P_\tau=$+1 for the $p^T_{\tau-jet}>$25 GeV cut (solid lines); and the decrease becomes 
steeper for the harder $p^T_{\tau-jet} > 50$~GeV cut (dashed lines) as expected. Thus 
this fractional cross section 
can be used to measure $P_\tau$.
In case the R $< 0.2$ region is inaccessible due to the difficulty in $\tau$ 
identification for a soft 
charged track, the cross section $\sigma(0.2 < R )$ can be used for normalization in 
the denominator of 
eq.~\ref{eq:twenty2}. To get a rough estimate of the size of uncertainty in the measurement of 
$P_\tau$, we have shown the predictions using the {\tt ALEPH} and 
{\tt CLEO}
parameterizations of $\tau$ decay~\cite{tauola} by two lines for each case. The horizontal spread 
between them corresponds to a 
$\Delta P_\tau = \pm 0.03$
($\pm$0.05) near $P_\tau$=$-$1($+$1). 
One should of course add to this the uncertainty in the measurement of f (eq.~\ref{eq:twenty2}). 
However being a ratio
of measured cross-sections this quantity is dominated by the statistical error, which 
is $\sim$0.01 corresponding to $\sim 10^4$ signal events. Hence the dominant error in 
$P_\tau$ measurement is expected to come from the parametrization of $\tau$ decay, as 
represented by the horizontal spread between the two lines of fig.2. It should be noted
here that even a conservative estimate of  
$\Delta P_\tau =\pm$0.1 would imply that the above four models can be distinguished 
from one another at the 5$\sigma$ level.

\section*{B) Partial Determination of SUSY Parameters:} In
this section we shall assume the mass relation of eq.\ref{eq:two} between the two
electroweak gauginos, which holds in a fairly broad class of SUSY models.
It holds in most SUSY models satisfying GUT symmetry, in which case 
the gaugino masses arise from the vev of a GUT singlet chiral superfield. 
This includes the SUGRA models with nonuniversal scalar masses. It holds for the GMSB 
as well. We shall assume that the $\tan\beta$ parameter can be independently measured
from the Higgs sector at LHC or ILC. Moreover we shall assume that the absolute scale of 
one of the SUSY masses, representing some combination of $M_1$, $M_2$ and $\mu$ can also 
be measured independently. 
Then the relative magnitude of
the $M_1$ and the $\mu$ parameters can be estimated from the $\tau$
Polarization.

We have calculated the composition of $\tilde Z_1$, as a function
of the ratio $M_1/|\mu|$, assuming the $\tilde Z_1$ mass of  
as in eq.\ref{eq:sixteen} to fix the SUSY mass scale. Fig. 3 shows the 
resulting $P_\tau$ from eqs.(\ref{eq:nine},\ref{eq:ten}) for two representative values 
of $\tan\beta = 40$ and 10 and $\cos\theta_{\tilde\tau}=$ 0.5 and 0.2.
It shows that $P_\tau$ is insensitive to this ratio for $M_1/|\mu|<$0.5 at $\tan\beta$=40
(b,d) and  $M_1/|\mu|<$1 at  $\tan\beta$=10(a,c). But above these limits one can measure 
this ratio to an accuracy of $\sim$
10\% with a $\Delta P_\tau=$0.1.

\section*{C) Complete determination of $EW$ SUSY parameters in
association with chargino $(\tilde W_1)$ mass and cross section measurements:}

In most of the SUSY benchmark points of Ref.~\cite{snowmass} discussed above, 
the lighter chargino $\tilde W_1$ mass lies in the range of 200-250 GeV. 
Therefore one hopes to see pair production of chargino in the first phase of 
ILC with a CM energy of $\sim$500 GeV. In that case one can estimate some of 
the above mentioned SUSY parameters from this process, as discussed 
in ~\cite{choi}. In particular the mass and the composition of $\tilde W_1$ 
can be easily obtained in terms of $M_2, \mu$ and $\tan\beta$ by diagonalizing 
the chargino mass matrix,
\br
{\cal M}_{\tilde c}=\left(\begin{array}{cc}
M_2 & {\sqrt 2} M_W \sin\beta  \\
{\sqrt 2} M_W \cos\beta & \mu
\end{array}\right)\qquad
\er
Thus one knows the coupling of a $\tilde W_1$ pair to the Z boson in terms of 
these three parameters, while it has the universal EM coupling to $\gamma$. 
With a right polarized electron beam the $\tilde W_1$ pair production proceeds 
only through s-channel $\gamma$ and $Z$ exchanges. Thus by measuring this 
production cross section $\sigma_R$ along with the $\tilde W_1$ mass from a 
threshold scan, one can estimate the composition of $\tilde W_1$. These two 
measurements determine two of the above three parameters.  Hopefully the third 
parameter ($\tan\beta$) can be independently estimated from the Higgs sector 
at the LHC or the ILC, as mentioned above.  Alternatively, one can measure 
the $\tilde W_1$ pair production cross sections with left and transversally 
polarized electron beams, $\sigma_L$ and $\sigma_T$. These two cross sections
get contributions from t-channel $\tilde\nu_e$ exchange along with the 
s-channel $\gamma$ and $Z$ exchanges.  Thus they depend on $\tilde\nu_e$ 
mass along with the above three parameters.  Nonetheless one can combine the 
three cross sections $\sigma_L, \sigma_R$ and $\sigma_T$ along with the 
$\tilde W_1$ mass measurement to determine all the four parameters, as 
discussed in ~\cite{choi}.  That would leave one SUSY parameter from the EW 
sector, $(M_1)$, which cannot be directly measured at ILC in the 
absence of a $\tilde Z_1$ pair production signal. However one can use 
instead the $P_\tau$ from the $\ta1 \to \tau \tilde Z_1$ decay to determine 
this parameter.  Recall that along with $M_1$ $P_\tau$ has a dependence
also on the stau sector mixing angle and $\tan \beta$. As a matter of fact, the
role of $P_\tau$ to probe the Higgsino component in the $\tilde Z_1$, especially
at large $\tan\beta$, has been emphasized\cite{boos}. As mentioned above, 
in the discussion we are assuming that these two parameters may be determined 
from other sectors. Of course, in a global fit, $P_\tau$ will be one of the 
important variables in a model independent determination of 
$\theta_\tau, M_2, M_1, \mu$ and $\tan \beta$.

Since both $M_2$ and $|\mu|$ would be known from chargino production, the 
$\tau$ polarization can be used to determine $M_1$ relative to either one of 
them. For illustrative purpose we have assumed $M_2=250$~GeV and three values 
of the ratio $|\mu|/M_2=$2, 1 and 0.5.  As in case of Fig.3 in this 
illustration also, we take two widely different sets of 
$\tan \beta, \cos\theta_\tau$, chosen to span the range of expectations for 
these two, to demonstrate the dependence of $P_\tau$ on them. 
The resulting polarization($P_\tau$) is shown against $M_1/M_2$
in Fig.4 for $\tan\beta=40$, $\cos\theta_{\tau}$=0.5 and in Fig.5 for  
$\tan\beta$=10, $\cos\theta_{\tau}$=0.2. One sees from these figures that while 
$P_\tau$ is flat for extreme values of $M_1/M_2$, it is a sensitive function 
of this ratio for $0.5 \lsim M_1/M_2 \lsim 2$. One can measure the 
ratio $M_1/M_2$ to an accuracy of 5-10\% over this interval with 
a $\Delta P_\tau=$0.1.  Thus knowing both $\mu$ and $M_2$ from 
$\tilde W_1$ production one can use $P_\tau$ to determine $M_1$ 
relative to $M_2$. In other words one can combine the $\tilde W_1$ mass 
and cross section measurements with the $\tau$ polarization to make a 
complete determination of the EW SUSY parameters.

It should be mentioned here that there is an alternative method of 
determining $M_1$ from kinematic distributions as suggested in 
ref.~\cite{zerwas}. We are not in a position to make a comparative assessment 
of the relative merits of the two methods. Instead we would like to emphasize 
their complementarity. The kinematic distributions probe the LSP($\tilde Z_1$) 
mass, while the $P_\tau$ measurement probes its composition. In this respect 
$P_\tau$ plays a role analogous to the production cross section of chargino
($\tilde W_1$). Note that one can combine the measurements of $\tilde W_1$ 
mass and composition via threshold scan and the production cross section 
$\sigma_R$ with those of $\tilde Z_1$ mass and composition via the kinematic 
distribution method and $P_\tau$. Thereby one can 
determine all the four parameters ($\mu, M_1,M_2$ and $\tan\beta$) with only the right 
polarized electron beam. On the other hand one can combine these four measurements with 
$\sigma_L$ and $\sigma_T$. Then the resulting redundancy of information can be used as a 
consistency 
check of the CP conserving MSSM. Alternatively, it can be used to extend the analysis 
and probe for CP violation in MSSM~\cite{kraml}.   

\section*{Acknowledgment:} The authors MG and DPR acknowledge CERN TH division for 
its hospitality during the final phase of this work.

\begin{figure}[htb]
\includegraphics[width=7in, height=4.in]{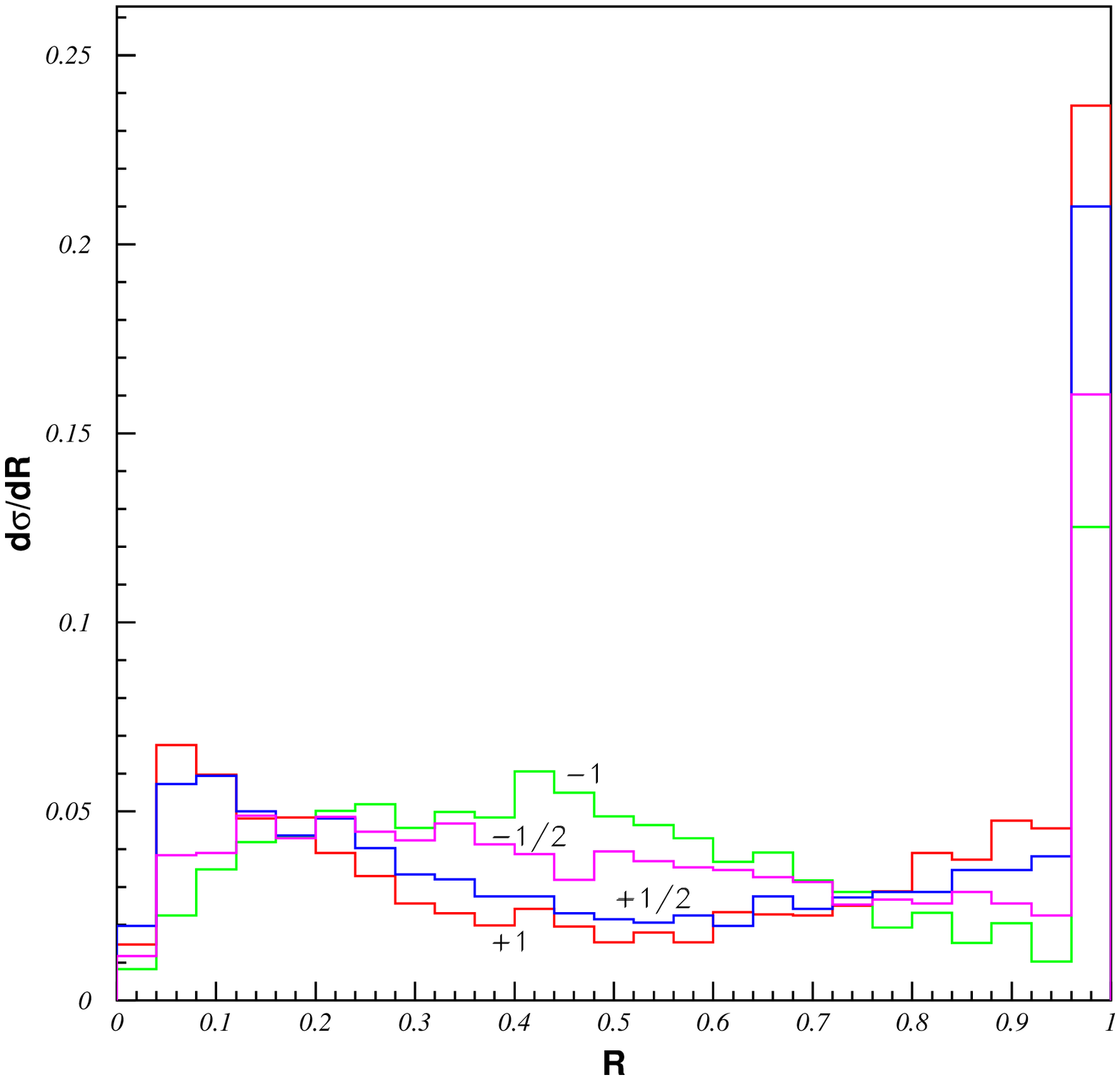}
\hspace{2cm}\includegraphics[width=7in, height=4.in]{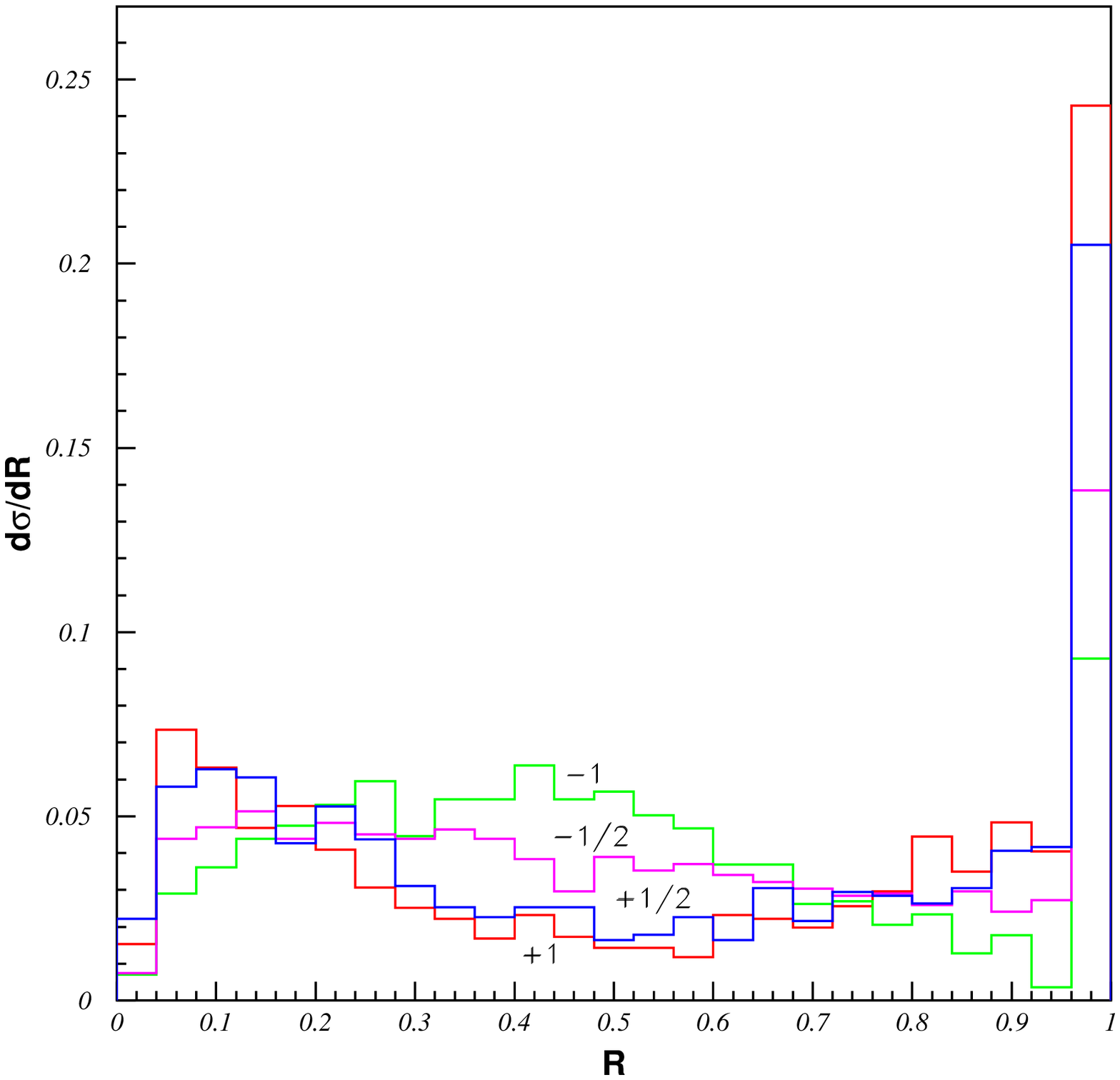}
\caption
{The normalized distributions in the fraction of $\tau$-jet momentum carried by the 
charged track for $P_\tau$=
+1, +1/2, -1/2 and -1. The upper and lower panels correspond to $p^T_{\tau-jet} >$25 
GeV and $p^T_{\tau-jet} >$50 GeV respectively.
}
\end{figure}
\begin{figure}[htbp]
\begin{center}
\hskip-1cm\centerline{\epsfig{file=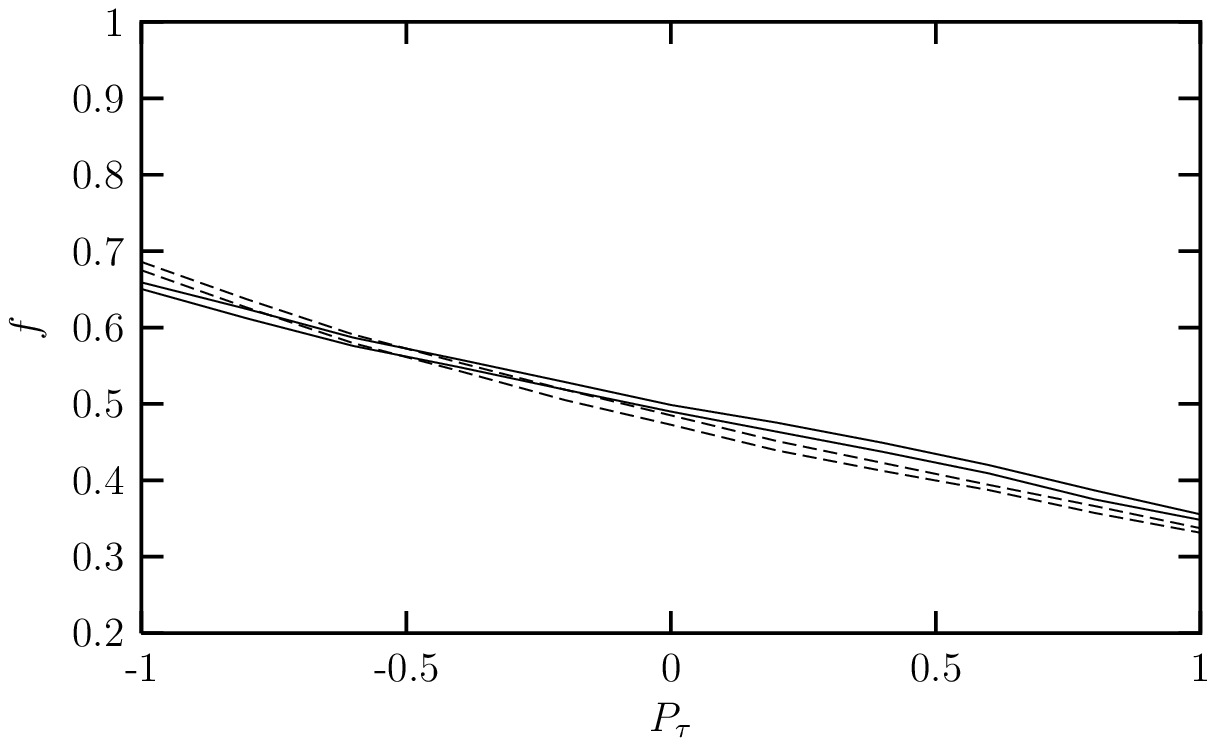,width=23cm}}
\vskip-15cm
\end{center}
\caption{
The fractional cross sections in the interval $0.2 < R < 0.8$ (eq.~\ref{eq:twenty2}) 
plotted 
against $\tau$ polarization for $p^T_{\tau-jet} >$25 GeV (solid lines) and 
$p^T_{\tau-jet} >$50 GeV (dashed lines). The upper and lower lines for each case 
correspond to the parameterizations of $\tau$ decay by the {\tt ALEPH} 
and the {\tt CLEO} collaborations~\cite{tauola}
respectively. 
}
\end{figure}
\begin{figure}[htbp]
\begin{center}
\hskip-1cm\centerline{\epsfig{file=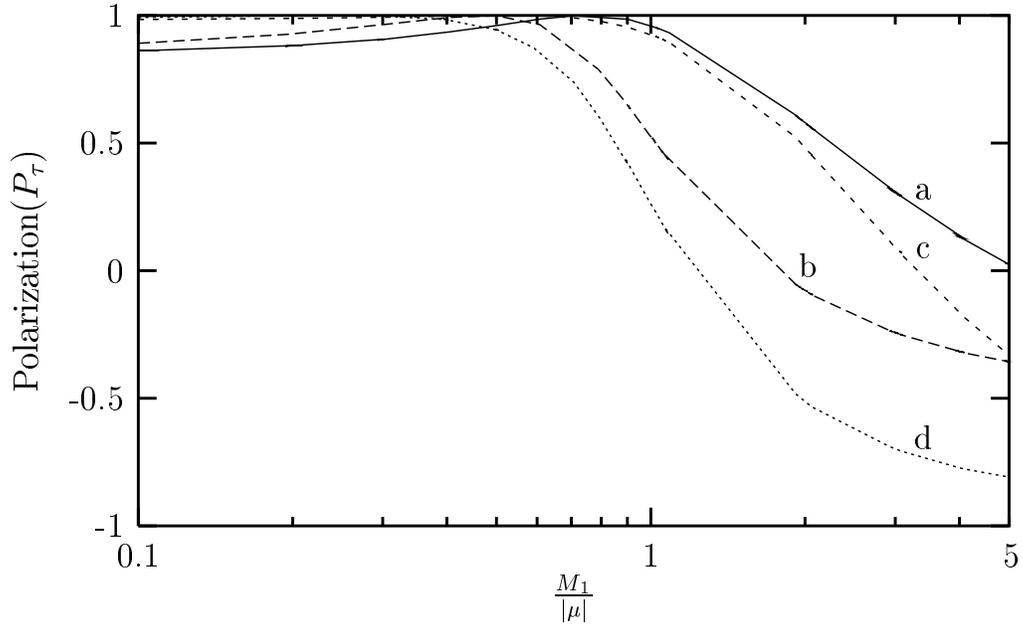,width=23cm}}
\vskip-15cm
\end{center}
\caption{
Polarization of $\tau$ shown against $M_1/|\mu|$ for a fixed value of 
$m_{\tilde Z_1} = 100$~GeV and $M_1/M_2$=0.5 and different
sets of parameters:(a)$\tan\beta=10, cos\theta_{\tau}=$0.5;
(b)$\tan\beta=40, cos\theta_{\tilde\tau}=$0.5;
(c)$\tan\beta=10, cos\theta_{\tilde\tau}=$0.2;
(d)$\tan\beta=40, cos\theta_{\tilde\tau}=$0.2.
}
\end{figure}
\begin{figure}[htbp]
\begin{center}
\vskip-5cm
\hskip-1cm\centerline{\epsfig{file=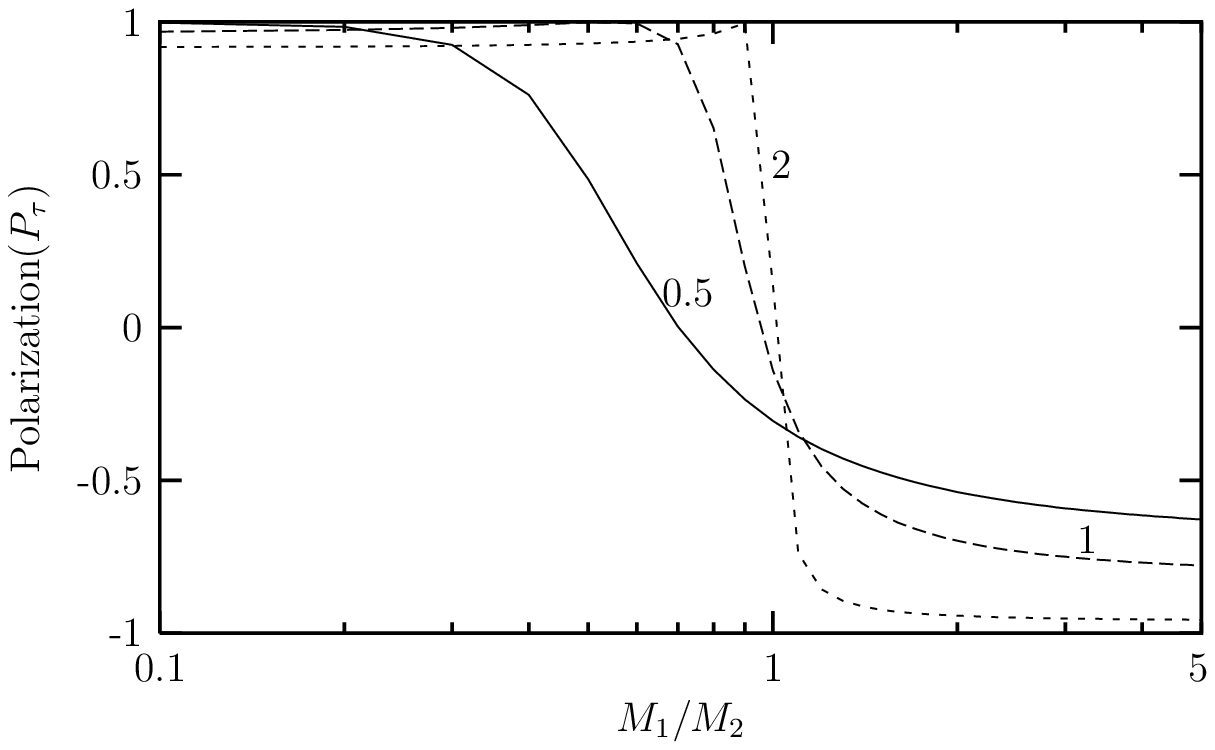,width=23cm}}
\vskip-15cm
\end{center}
\caption
{Polarization of $\tau$ shown against $M_1/M_2$ for a fixed value of 
$\tan\beta=40, cos\theta_{\tau}=$0.5, $M_2=$250~GeV 
and $|\mu|/M_2=$0.5,1,2.
}
\end{figure}
\begin{figure}[htbp]
\begin{center}
\vskip-5cm
\hskip-1cm\centerline{\epsfig{file=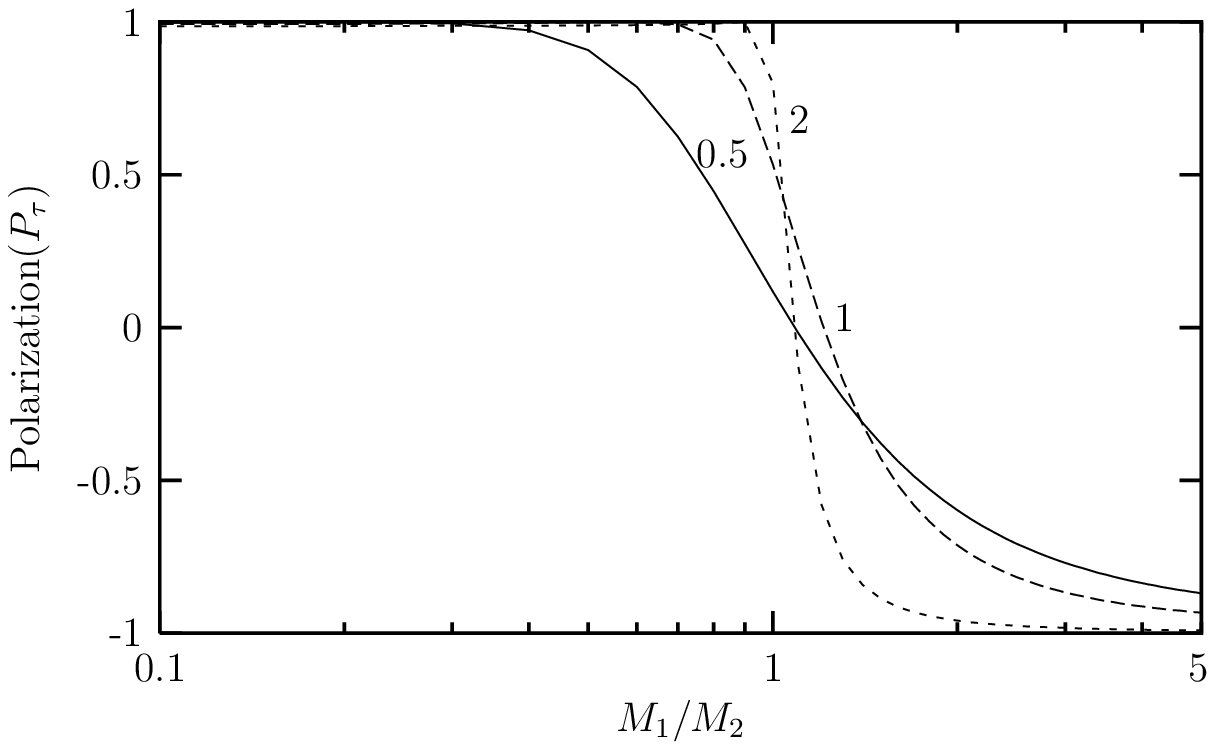,width=23cm}}
\vskip-15cm
\end{center}
\caption
{Same as in Fig.4, but for $\tan\beta=10, cos\theta_{\tau}=$0.2.
}
\end{figure}

\end{document}